\documentclass[twocolumn,twoside,slac_two]{revtex4}
\usepackage{graphicx}
\usepackage{fancyhdr}
\usepackage{hyperref}
\hypersetup{
    colorlinks=true,
    urlcolor=magenta,
}
\begin{document}

\title{Improving reconstruction methods for radio
measurements with Tunka-Rex}

%

\author{P.A. Bezyazeekov\textsuperscript{a}, N.M. Budnev\textsuperscript{a}, O. Fedorov\textsuperscript{a}, O.A. Gress\textsuperscript{a}, A. Haungs\textsuperscript{b}, R. Hiller\textsuperscript{b}, T. Huege\textsuperscript{b}, 
Y. Kazarina\textsuperscript{a}, M. Kleifges\textsuperscript{c}, E.E. Korosteleva\textsuperscript{d},
D. Kostunin\textsuperscript{b}, O. Kr\"omer\textsuperscript{c}, V. Kungel\textsuperscript{b}, L.A.~Kuzmichev\textsuperscript{d}, V. Lenok\textsuperscript{a}, N. Lubsandorzhiev\textsuperscript{d}, T. N. Marshalkina\textsuperscript{a}, R.R. Mirgazov\textsuperscript{a}, R. Monkhoev\textsuperscript{a}, E.A. Osipova\textsuperscript{d},
A. Pakhorukov\textsuperscript{a}, L. Pankov\textsuperscript{a}, V.V. Prosin\textsuperscript{d}, F.G. Schr\"oder\textsuperscript{b}, A. Zagorodnikov\textsuperscript{a}
}

\affiliation{
\textsuperscript{a}Institute of Applied Physics ISU, Irkutsk, Russia;
}
\affiliation{
\textsuperscript{b}Institut f\"ur Kernphysik, Karlsruhe Institute of Technology (KIT), Germany;
}
\affiliation{
\textsuperscript{c}Institut f\"ur Prozessdatenverarbeitung und Elektronik, KIT, Germany;
}
\affiliation{
\textsuperscript{d}Skobeltsyn Institute of Nuclear Physics MSU, Moscow, Russia;
}

\begin{abstract}Tunka-Rex is a detector for radio emission produced by cosmic-ray air showers. 
It is located in Siberia, and triggered by Tunka-133, a co-located air-Cherenkov detector during night, and by the scintillator array Tunka-Grande during day.
Tunka-Rex demonstrates that the radio technique can provide a cost-effective extension of existing air-shower arrays.
Operating in the frequency range of 30-80 MHz, Tunka-Rex is limited by the galactic background and suffers from the local radio interferences.
We investigate the possibilities of improving measured data using different approaches, particularly, a multivariate treatment of background is considered, as well as an improved likelihood fit of the lateral distribution of amplitudes.
\end{abstract}

\maketitle

\thispagestyle{fancy}


\section{Introduction}
Tunka-Rex~\cite{TunkaRex_NIM_2015,Kostnin:2017bzd} is an antenna array for the detection of radio emission of extensive atmospheric showers (EAS) created by cosmic rays~\cite{Schroder:2016hrv}.
It works jointly with the non-imaging air-Cherenkov light detector Tunka-133~\cite{Prosin:2014dxa} and the scintillators of Tunka-Grande~\cite{Budnev:2015cha} and receives triggers from both of them.
Comparing to classical optical methods of EAS detection, detection of radio emission is interesting because it is independent of the time of the day (measurements are possible during sunlight) and of weather conditions (except thunderstorms).
Consequently, radio has higher duty-cycle than optical methods.
Tunka-Rex antennas are stable and cost-effective and able to operate in a sparse large-scale configuration~\cite{Bezyazeekov:2015ica,Apel:2016gws}.

At the present moment Tunka-Rex consists of 63 antennas covering an area of about 3 km\textsuperscript{2}. 
Each Tunka-Rex antenna station consists of two perpendicularly 
aligned SALLAs (Short Aperiodic Loaded Loop Antenna)~\cite{KroemerSALLAIcrc2009,Abreu:2012pi} with 120~cm diameter mounted on a wooden pole on a height of about 2.5~m.
At the top they are connected to a low noise amplifier (LNA).
This LNA is connected to a filter-amplifier via 30 m coaxial RG213 cables. 
THe filter-amplifier is connected to the Tunka-133 or Tunka-Grande FADCs with a short 1 m RG52 cable. 
The sampling rate of the FADCs is 200 MHz, and the length of the trace is 1024 samples with a bit depth of 12 bit.

Tunka-Rex operates in an environment with low signal-to-noise ratio (SNR) conditions, that requires a sophisticated post-processing of recorded traces.
The standard approach is to use different filters (broadband or narrowband) to increase the SNR.
Then, the signal peak is defined as maximum of the filtered trace.
The aim of the present work is developing and testing new methods for the improvement of signal reconstruction and for fitting the lateral distribution of the radio amplitudes.

\section{Standard procedure of data reconstruction}
For signal reconstruction Tunka-Rex uses a modified radio extension of the Offline software framework developed by the Pierre Auger Collaboration~\cite{Abreu:2011fb}.
Raw Tunka-Rex data consist of ADC traces from the detector containing measured amplitudes in two orthogonal polarizations.
The signal reconstruction includes following steps:
\begin{enumerate}
\item Upsampling of recorded traces by a factor of 4, and applying the following filters: first, a band-stop filter suppresses narrowband interferences occurring each 5 MHz; second, a bandpass filter, restricts the band to 35--76 MHz;
\item The amplitude of the measured signal $S$ is defined as maximum of Hilbert envelope in the signal window of the recorded trace;
\item The noise level $N$ is defined as RMS of amplitudes in the noise window.
\end{enumerate}

The SNR is calculated as square of ratio between measured signal and noise:
\begin{equation}
\mathrm{SNR} = {S}^{2}/{N}^{2}\,.
\end{equation}
Antenna stations with $\mathrm{SNR} < 10$ in at least one of the channels are rejected.

The rest of the stations is used for the reconstruction of the arrival direction by fitting peak times with a plane wavefront model.
Knowing the arrival direction (shower axis), the electrical field is reconstructed by applying the antenna pattern and assuming that the electrical field along the shower axis is zero.
Then the cut ${\mathrm{SNR} \geq 10}$ is applied again for the vectorial traces of the electrical field on the antenna station.

Let us consider the influence of the noise on the measured signal.
The signal measured at the frequency $\nu$ is defined as sum of the true signal and the noise contribution:
\begin{equation}
A^\nu_{m}e^{i\phi_m} = A^\nu_{t}e^{i\phi_{t}} + A^\nu_{n}e^{i\phi_{n}},
\end{equation}
where $A^\nu_{m}$, $A^\nu_{t}$, $A^\nu_{n}$ are the amplitudes of the measured, true and noise signals, in the frequency domain,
and $\phi_m$, $\phi_{t}$, $\phi_{n}$ are their phases.
The unknown phase of the noise component can shift resulting amplitude by $\pm$30\% for ${\mathrm{SNR} = 10}$.

\begin{figure}[t]
\includegraphics[angle=270,width=1.0\linewidth]{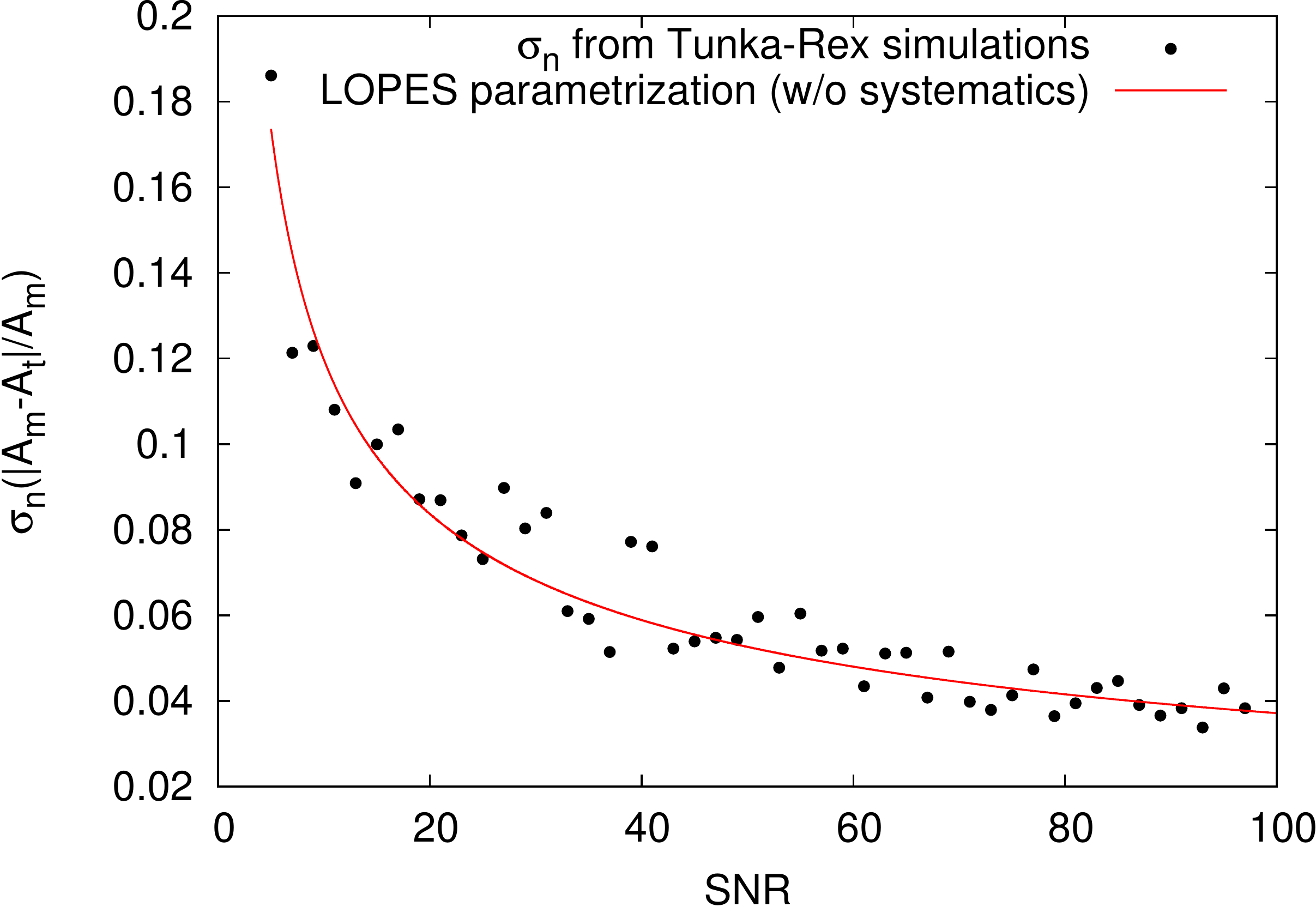} 
\caption{Dependence of the uncertainty of the signal reconstruction on the SNR.
The points are obtained as standard deviation of normalized difference between simulated and reconstructed amplitudes using the Tunka-Rex noise library and hardware response (the spread between neightboring bins is due to low statistics);
the curve is obtained from a LOPES parameterization (see Eq.~\ref{eq_s}), and is in agreement with Tunka-Rex simulations.}
\label{fig:amp_uncert}
\end{figure}

The contribution of the noise to the total power of measured signal is corrected using the parameterization:
\begin{equation}
A_{t} = f(\mathrm{SNR})A_{m}\,,
\end{equation}
where
\begin{equation}
f(\mathrm{SNR}) = \sqrt{1-k/\mathrm{SNR}}\,,
\end{equation}
where parameter $k$ is extracted from a fit to simulations~(details in \cite{Bezyazeekov:2015ica}), and then this parameterization is applied to the measured signals.
Besides this effect, the measured amplitude is characterized by the uncertainty defined as standard deviation $\sigma_n$ of $(A_m - A_t)/A_m$.
This uncertainty is parameterized by LOPES~\cite{Schroeder2012S238} as a function of $\mathrm{SNR}$ (see Fig.~\ref{fig:amp_uncert}):
\begin{equation}
\sigma_n = \frac{a}{\sqrt{\mathrm{SNR} - a^2}} \left(b + c \exp (-\sqrt{\mathrm{SNR}}/a)\right)\,,
\label{eq_s}
\end{equation}
where ${a = 0.602}$, ${b = 0.616}$ and ${c = 0.213}$ are dimensionless parameters. 
This uncertainty is included in the chi-square fit of the lateral distribution function (LDF)~\cite{Kostunin:2015taa}:
\begin{equation}
\mathcal{E}(r)=\mathcal{E}_{0}\exp\left(a_{1}(r-r_{0})+a_{2}(r-r_{0})^{2}\right)\,,
\end{equation}
where $\mathcal{E}(r)$ is the amplitude at the antenna station with distance $r$ from the shower axis, 
$\mathcal{E}_{0}$ is a normalization factor proportional to the energy of the primary cosmic ray, 
$r_{0}$ is an arbitrary parameter chosen to obtain maximum correlation with the primary energy and the distance to the shower maximum, 
$a_{1}$ and $a_{2}$ are parameters proportional to the slope and width of the LDF, respectively.

\section{Signal reconstruction with a neural network}
\begin{figure}[t]
\includegraphics[width=1.0\linewidth]{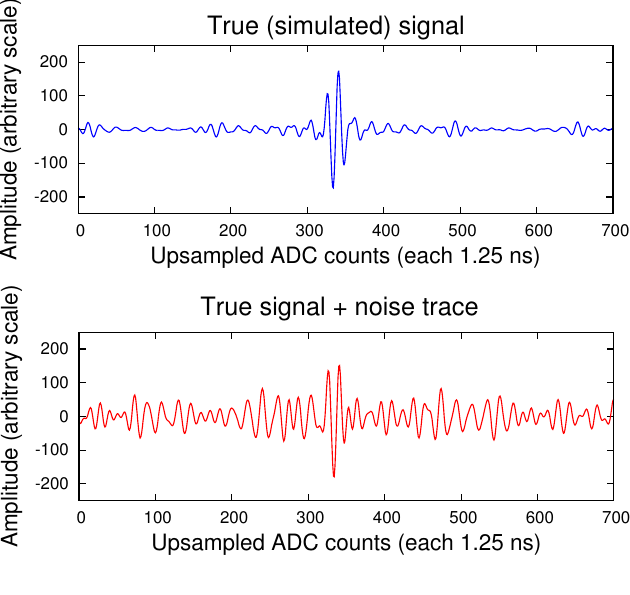} 
\caption{Example of ADC trace with clear (top) and distorted (bottom) signal.
}
\label{fig:trace}
\end{figure}

For the design of the neural network we use the PyBrain library~\cite{pybrain2010jmlr} written in Python.
At the first stage we have trained a test network which performs the classification of the transient background in a signal window.
It is simple ``perceptron'' (neural network with single hidden layer) trained on a dataset containing separated samples containing pure noise and samples with background pulses
The network returns two values: first the probability of pure noise and, second, the probability of background pulse (with high SNR). 
Resulting efficiency of that classification is 98\%.
For the amplitude reconstruction we designed another type of neural network, with the following structure:
\begin{itemize}
\item 200 input neurons
\item 3 hidden layers, with each 500 neurons
\item 1 output neuron
\end{itemize}
The activation function of the neurons is sigmoid, for training we use resilient propagation method. 
The input layer takes the Hilbert envelope of trace with 200 ADC counts of the trace centered around the assumed signal as input.
The output neuron returns the reconstructed amplitude.
As training dataset we simulated radio pulses with CoREAS~\cite{Huege:2013vt} (for two Tunka-Rex channels) and then distort them using a background library collected by the Tunka-Rex experiment.
The example of traces is given in Fig.~\ref{fig:trace}.
Training dataset contains about 7000 distorted traces with different SNR levels and known amplitude of true (simulated) signal.
In each training epoch, the network takes 200 values of distorted traces as input and the value of the true amplitude as output.

For training and test of the neural network we selected only traces containing a true signal higher than the noise level (i.e. ${\mathrm{SNR}>1}$).
Thus, network is not able to distinguish between pure noise and low-SNR signals, but is able to reconstruct amplitude for signals with ${\mathrm{SNR} > 1}$.

\begin{figure}[t]
\includegraphics[width=1.0\linewidth]{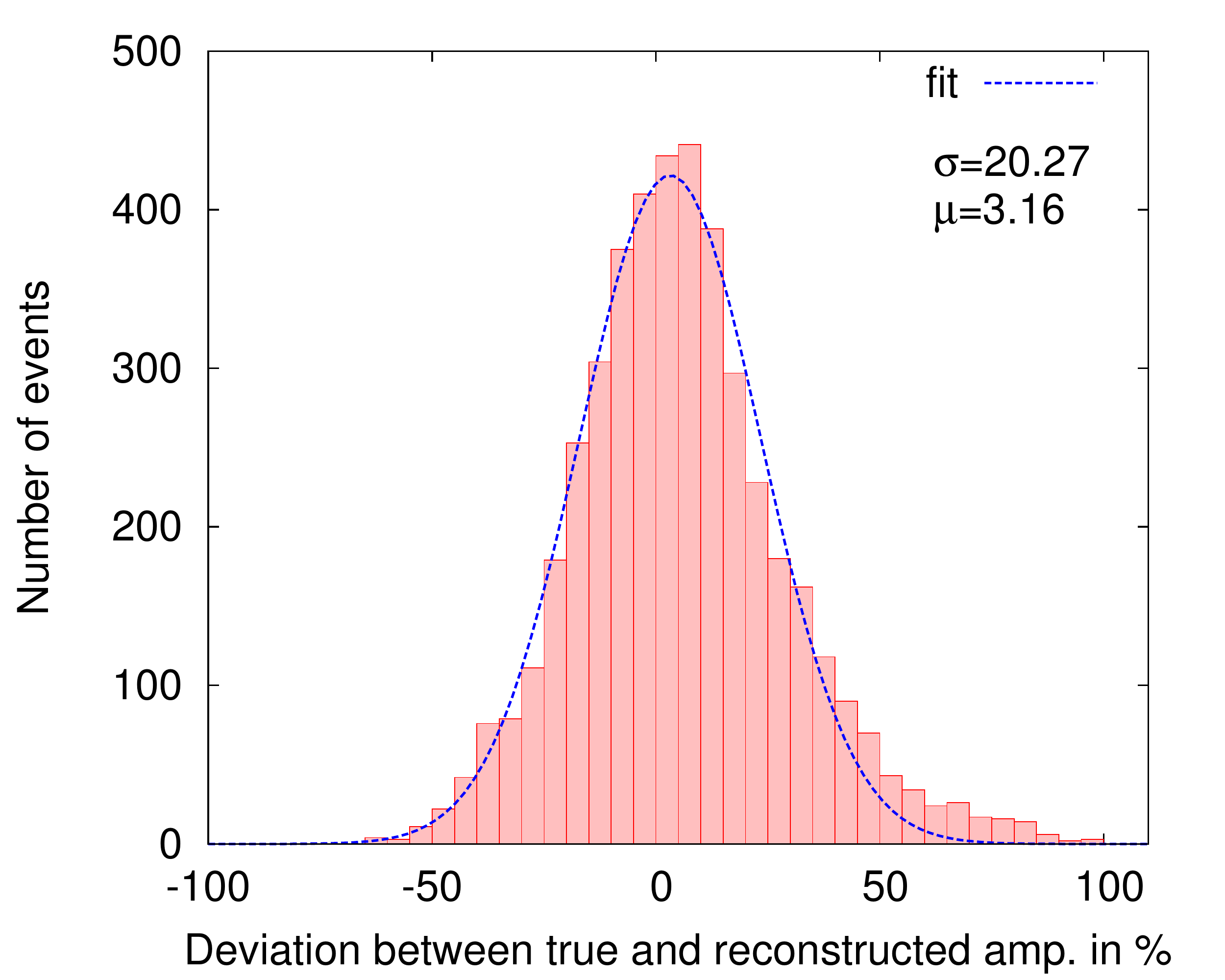} 
\caption{Deviation between true and reconstructed amplitude given by the neural network for signals with ${\mathrm{SNR} < 5}$.
}
\label{1}
\end{figure}

Running the designed neural network on the testing dataset has shown that the uncertainty of the amplitude reconstruction is about 20\%, which is worse, than given by our standard methods.
However, the same uncertainty is obtained for very low SNRs, and the correction for the bias due to noise is obtained automatically.
The distribution of $(A_m - A_t)/A_m$ for signals with ${\mathrm{SNR} < 5}$ processed with the neural network is given in Fig.~\ref{1}.

\section{Optimization of likelihood LDF fit}

\begin{figure}[t]
\includegraphics[width=0.8\linewidth]{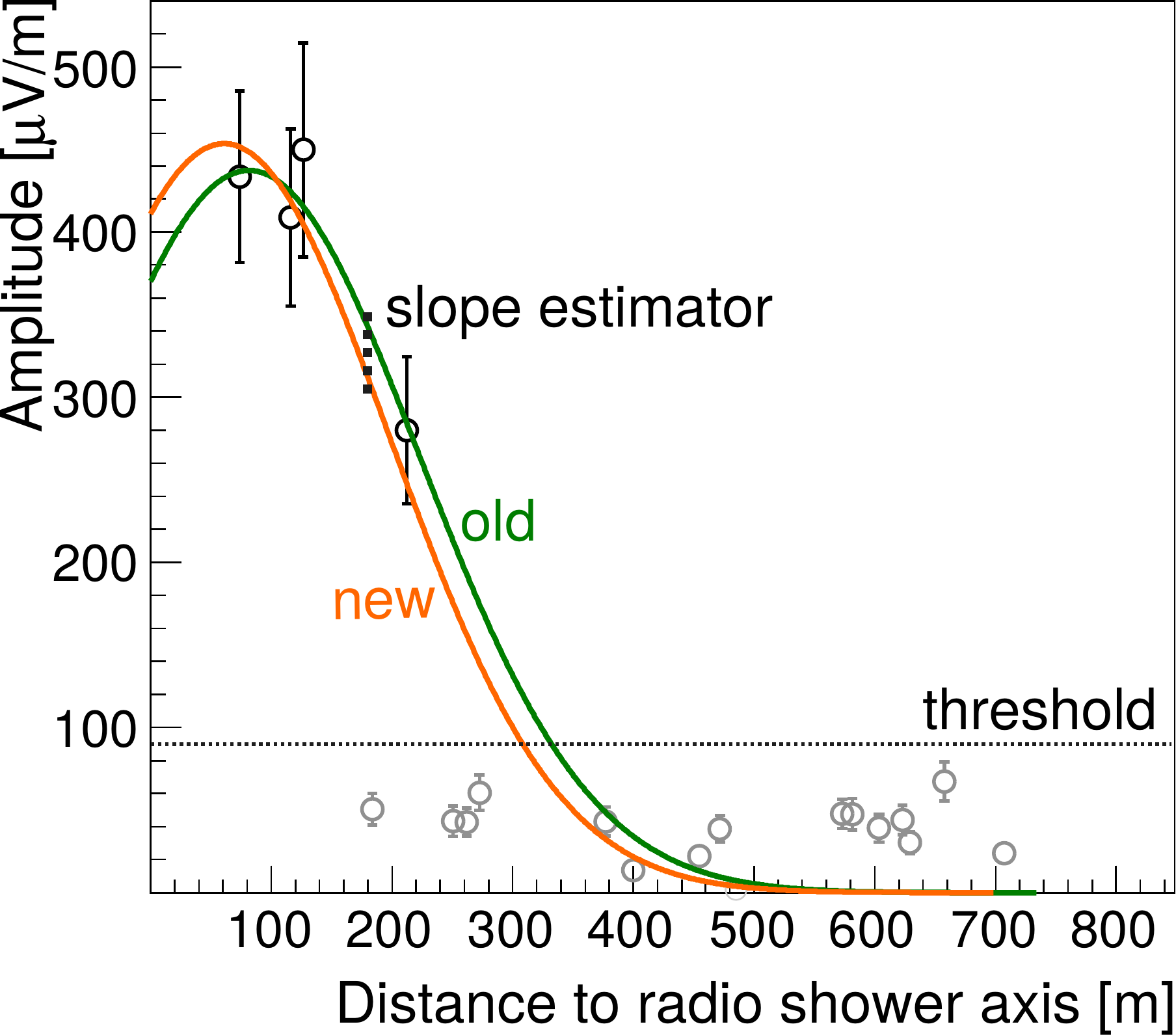} 
\caption{Difference between best LDF fits using standard (old) and modified (new) likelihood functions.
Although the visual change of the slope is not very significant, the reconstructed depth of shower maximum has changed by about 100~g/cm\textsuperscript{2} in this case.
Let us note, that such events are not passing standard $X_\mathrm{max}$ quality cuts of Tunka-Rex reconstruction.
}
\label{fig:ldf_newfit}
\end{figure}

For the fitting of lateral distribution function (LDF) the standard chi-square fit is used, where uncertainties are defined as a function of SNR (see Sec.~2).
This way, the amplitudes from antennas stations with ${\mathrm{SNR} > 10}$ are included in the fit and weighted with their uncertainties.
The information from the antenna stations without signal was not used.
However, they still contain information about the confidence levels of the signal amplitudes, i.e. these stations provide an upper limit for the amplitude.
They are weighted according to the general threshold of signal detection if the fitted LDF is higher than this threshold, and the get zero weigth if the LDF is already below the threshold.

Thus, we modified the LDF fit and defined a likelihood function as follows:
\begin{eqnarray}
&& \mathcal{L} = \prod_{i=0}^N \exp \bigg\{ - \bigg[ \frac{( f_i - y_i )^2}{2 \sigma_i^2} \Theta(y_i - y_{th}) \nonumber \\ 
&& +\frac{( f_i - y_{th} )^2}{2 y_{th}^2} \Theta(f_i - y_{th}) \Theta(y_{th} - y_i) \bigg] \bigg\}\,,\\
&&\mbox{and the chi-square becomes} \nonumber \\
&& \chi^2 = \sum_{i=0}^N \bigg[ \frac{( f_i - y_i )^2}{\sigma_i^2} \Theta(y_i - y_{th}) \nonumber \\
&& +\frac{( f_i - y_{th} )^2}{y_{th}^2} \Theta(f_i - y_{th}) \Theta(y_{th} - y_i) \bigg]\,,
\end{eqnarray}
where $f_i$ and $y_i$ are the fitted and measured values of the amplitudes, $\sigma_i$ is the uncertainty of $i$-th station of $N$, $\Theta(x)$ is the Heaviside step function.

One can note, that antenna stations with amplitudes ${y_i < y_{th}}$ are weighted with $y_{th}$, where ${y_{th} \approx 90}$~{$\mu$}V/m is the threshold of signal detection~\cite{HillerArena2016}.
Since ${y_{th} \ge 10\sigma_i}$ for $\mathrm{SNR} > 10$ the weights of the antenna stations without signals are relatively small.
The first tests of this modified LDF fitting have shown, that the reconstruction of air-shower parameters is slightly improved, however this simple parameter-free weighting can be further developed.
An example of a modified fit is given in Fig.~\ref{fig:ldf_newfit}.
It is worth noticing, that only reconstruction of low-quality events (e.g. not passing $X_\mathrm{max}$ quality cut, see Ref~\cite{Bezyazeekov:2015ica} for details) is significantly affected, the result of the reconstruction of the standard event set is almost the same.
Which means, that such improvement of LDF fitting can increase the statistics near threshold.

\section{Summary}
We present our current progress on the improvements of reconstructing radio events from cosmic-ray air-showers.
We have worked in two directions: reconstruction of the low-SNR radio pulses using neural networks, and improving the fit of the LDF by including stations with amplitudes below the SNR cut.
\begin{itemize}
\item \textit{Low-SNR signal reconstruction}.
This study shows that machine learning has a potential for solving tasks related to multi-variable conditions, like radio background.
For high SNRs the resulting precision of the reconstructed amplitude is worse comparing to our standard methods, however the neural network shows better performance at low SNRs~${(<5)}$.
We plan to continue this study by testing different configurations of neural networks and another machine learning methods, e.g decision trees.
\item \textit{Modification of LDF fit}.
We added weights of antenna stations without signal into the standard chi-square fit.
The simple parameterization of the weights shows a slight improvement of the air-shower reconstruction.
We plan to tune parameters in order to optimize weighting of the antenna stations without signal.
After this we include this stations also for shower core reconstruction with radio.
\end{itemize}

\section*{Acknowledgments}
The construction of Tunka-Rex was funded by the German Helmholtz Association and the Russian Foundation for Basic Research (grant HRJRG-303). Moreover, this work was supported by the Helmholtz Alliance for Astroparticle Physics (HAP), by Deutsche Forschungsgemeinschaft (DFG) grant SCHR 1480/1-1, and by the Russian Federation Ministry of Education and Science (agreement 14.B25.31.0010). Moreover, this work was supported by the Russian Fund of Basic research grants 16-32-00329 and 16-02-00738.

\bibliography{references}

\end{document}